\documentstyle[11pt,newpasp,twoside,epsf]{article}
\def\edcomment#1{\iffalse\marginpar{\raggedright\sl#1\/}\else\relax\fi}
\reversemarginpar

\begin{document}
\title{Gamma-Ray Burst Detection with INTEGRAL/SPI}
 \author{Andreas von Kienlin, Nikolas Arend, Giselher Lichti, Andrew Strong}
\affil{Max-Planck-Institut f\"ur extraterrestrische Physik, Garching, Germany}
 \author{Paul Connell}
\affil{University of Birmingham, Edgbaston, Birmingham, United Kingdom}

\begin{abstract}
The spectrometer SPI, one of the two main instruments of the INTEGRAL spacecraft, has strong capabilities in the 
field of Gamma-Ray Burst (GRB) detections. In its 16$^o$ field of view (FoV) SPI is able to trigger and to localize 
GRBs with an accuracy for strong bursts better than 1$^o$. The expected GRB detection rate is about one per month.
\end{abstract}

\section{Introduction}

The detection and investigation of cosmic gamma-ray bursts (GRBs) is one of the important scientific topics of ESA's 
INTEGRAL mission, which will be launched on Oct.\ 17, 2002. From all instruments of the INTEGRAL payload, the two 
main instruments, the spectrometer SPI and the imager IBIS, and the two monitors, JEM-X and OMC, contributions to the 
science of GRBs are expected. Especially the sensitivity of INTEGRAL's imager and spectrometer above 20 keV will 
distinguish them from other missions. Although INTEGRAL is not equipped with an on-board burst-alert system it has 
strong capabilities to detect bursts thanks to the ground-based INTEGRAL burst-alert system IBAS (Mereghetti 2001). 
IBAS is an automatic software system for the near-real time distribution of GRB alerts detected by INTEGRAL to the
scientific community. The telemetry 
stream of the satellite, which is linked via the mission operation center (MOC) at Darmstadt, Germany, to the INTEGRAL 
science data center (ISDC) at Versoix, Switzerland, will be monitored continuously for the occurrence of GRBs. The  expected 
transmission time of the telemetry from the satellite to ISDC  will be approximately 30 s. Additionally the processing time 
at ISDC  and the time needed to distribute the data to the clients has to be added. In total an observer could expect the 
first alert message within less than 1 min after the onset of the burst. Thus, in the case of a long-duration burst, follow-up 
observations are possible even during the active phase of the burst.

\section{Burst detection in SPI's field of view}

The aim of the spectrometer SPI is to perform high-resolution spectroscopy in the energy range between 20 keV and 8 MeV. 
The imaging capability will be good, but it is exceeded by that of the imager IBIS which complements SPI with its 
high imaging resolution, but with less spectroscopic resolving power. So the expected yield and discoveries in the field
of GRB research will be of a different kind for each of the four  INTEGRAL instruments. With SPI new spectroscopic discoveries 
are possible, supposing that previously-unkown spectral features exist. IBIS will 
provide more precise GRB locations to the community, which is important for the observation of GRB afterglows. The minor 
sensitivity of SPI for GRB detection is compensated by  its larger FoV. So the expected rate/year of detected GRBs is 
about the same for IBIS and SPI. 

The camera of SPI, which consists of 19 cooled high-purity germanium detectors, residing in a cryostat,
is shielded on the side walls and rear side by a large anticoincidence shield (ACS).
The field of view (FoV) of the camera is defined by the upper opening of the ACS.  A detailed description of the ACS and its capabilities 
for the burst detction can be found in (von Kienlin 2001) and (von Kienlin 2003). The imaging
capability of the instrument is attained by a passive-coded mask mounted 1.7 m above the camera. 
By observing  a series of nearby pointing directions around the source("dithering") the imaging capability of SPI is 
improved by reducing ambiguity effects.  
For the normal mode of operation two dithering patterns are planned, 
a quadratic $5 \times 5$ and a hexagonal with 7 pointings. The pointing on one grid point will last for about 20-30 min 
with a short 5 min slew between succeeding pointings. Due to the short duration of the bursts, dithering will not improve 
the imaging of SPI for bursts. But during its short duration a burst will be in most cases the brightest $\gamma$-ray source for SPI 
in the whole FoV,  so imaging with SPI is still possible. 
Despite the modest angular resolution of SPI, which is in the order of  2$^o$ , it is possible to locate the direction of 
bursts down to a few arcminutes. 

The triggering algorithm which will be used for the burst search in SPI's FoV will be
similiar to that used inside IBAS for the IBIS-ISGRI instrument.
The first algorithm is based on imaging only. It looks for the appearance of new sources by comparing the actual 
image of the sky with those obtained  previously. This algorithm is more sensitive to slowly rising bursts, since the imaging
routine needs some time. The second is expected to be faster, because it monitors the overall event rate in all detectors 
together as function of time. In the case of a trigger the second algorithm uses the imaging algorithm for 
verification.

Every interaction of a $\gamma$-ray in one of the Ge-detectors is downlinked event-by-event, wrapped into packets within the telemetry stream.  
The event-by-event packets contain information on the detector ID which was hit, the energy channel of the recorded pulse-height signal and the time of
interaction, with an accuracy of about 100 $\mu$s. Two kinds of interactions of $\gamma$-rays in the Ge-camera have to be distinguished, in the first case 
a $\gamma$-ray  only interacts within one of the Ge-detectors (single-detector events), in the second case the $\gamma$-ray is interacting with 
more than one detector (multiple-detector events). The condition for the occurrence of a multiple-detector event is
defined by a coincidence window inside the digital front-end electronics (DFEE) of SPI.
Additional information on the detector interaction is available through the science house keeping data (HK-data). The rate for each detector is recorded 
on a 1 s basis inside the DFEE and sent to ground, but without energy information. The advantage of these data is that they are always 
available, even in the case of telemetry limitations.  

Simulations and calculations have been performed (Skinner 1997) in order to investigate SPI's capabilities 
for the burst detection in its FoV. 
For bursts with a fluence of $1.1 \times 10^{-6}$~erg/cm$^2$ and a duration of 1 s (only accounting for single detector 
events in the energy range from 20 to 300 keV) the positioning errors inside the FoV ($< 9^o$) will be between 1.5' and 20' and for bursts 
with a larger offset ($< 15^o$) the position error will still be less  than $1^o$. 
Problems arising from ambiguities and systematic
errors in the determination of the position are more serious towards the edge of the FoV. In summary GRB source locations 
to ~10' are possible over a wide FoV. 
For weak bursts the detection could be problematic, the imperfect imaging together with random fluctuations can lead to a 
spurious peak appearing to be stronger than the true one. The best-estimate position can then be far from the correct location.
The ambiguity problem can occur for bursts with a strength of less than $10^{-7}$~erg/cm$^{2}$.
As described above, one method for detecting bursts in the SPI data is the monitoring of the overall
event rate in all detectors together as function of time.  
The minimum-size burst detectable for a backgound rate of ~0.25 cts cm$^{-2}$~s$^{-1}$ in the 50 to 300 keV regime for a 1 s burst
is about 0.2 photons (Skinner 1997), corresponding to $4~ \times 10^{-8}$~erg/cm$^{2}$. This is about the same as the weakest bursts 
detected by BATSE. 
The rate at which GRBs are detected by an instrument depends on the FoV as well as on the sensitivity. SPI can detect 
bursts at off-axis angles up to at least 15$^o$, so the useful FoV is quite large, ~ 0.2 sr or 680 square degrees 
and consequently the expected detection rate of about 1 burst per month is comparable for IBIS and SPI.

   \begin{figure}
   \plotone{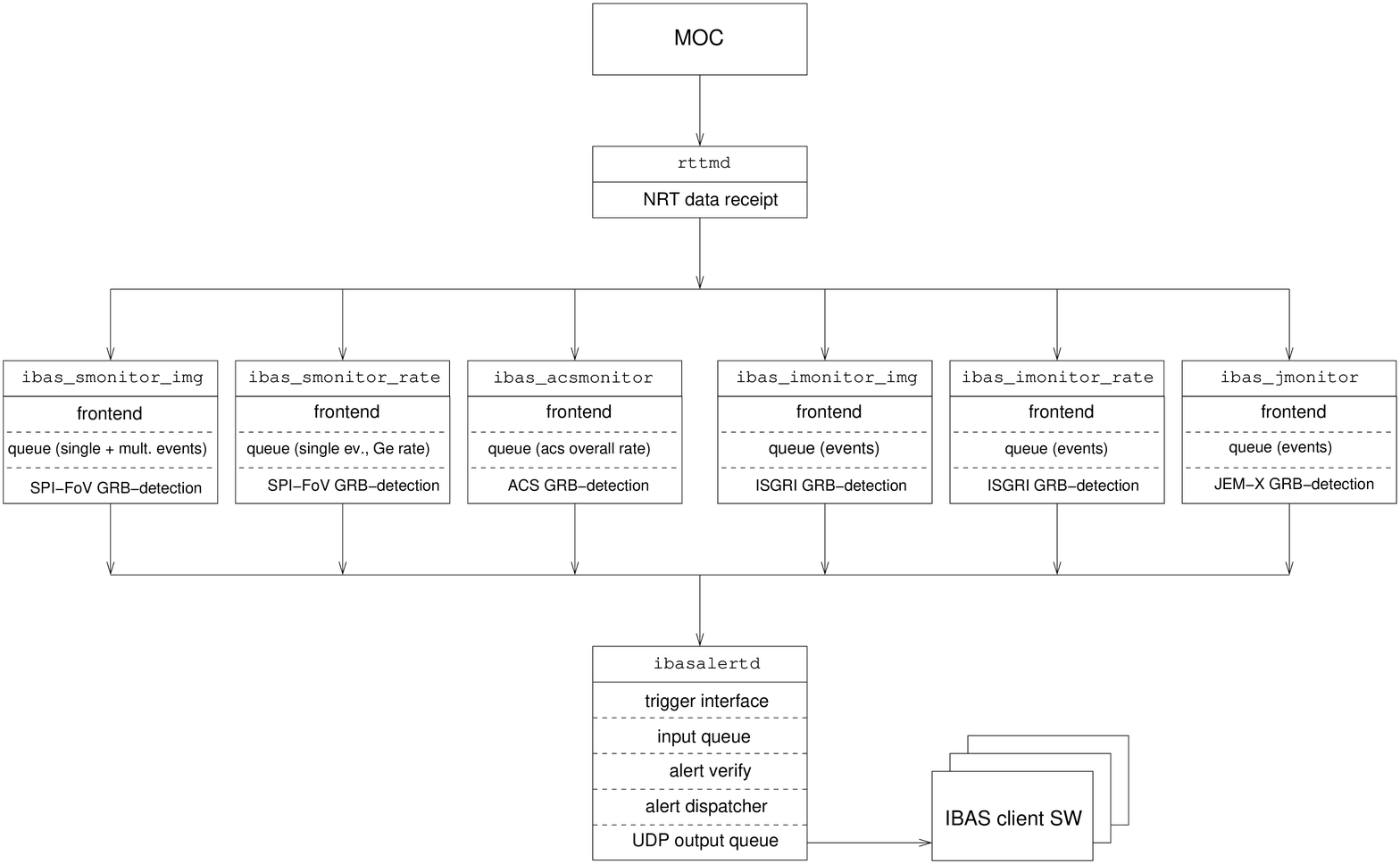}
    \caption{Structure of the INTEGRAL burst-alert system IBAS.}
   \end{figure} 

\section{GRB detection in SPI's FoV with IBAS}

Figure 1 gives a simplified overview on the structure of the whole IBAS system. After the near-real-time data receipt
of the telemetry from MOC, several GRB detection programs are searching in the telemetry stream for the occurrence of a burst. 
Three of them are responsible for the spectrometer SPI, there are the two programs monitoring the camera events 
({\tt ibas\_smonitor\_img} and {\tt ibas\_smonitor\_rate}) and the monitor program for the ACS overall veto rate ({\tt ibas\_acsmonitor}). 
The other half of the branches monitor the science data of IBIS ISGRI ({\tt ibas\_imonitor\_img} and {\tt ibas\_imonitor\_rate}) 
and JEMX ({\tt ibas\_jmonitor}). The program {\tt ibasalertd}   receives the GRB triggers from the individual branches and 
performs the final alert verification and distribution of the alert messages to subscribed users. 
All IBAS processes are multithreaded applications and run as daemon processes.  
So IBAS is able to perform several subtasks at the same time, with the advantage for the burst search to monitor the telemetry  in 
different energy bands and with different time binning. 
Each of the GRB detection processes uses the frontend library to read out 
the telemetry packets provided by the data receipt. Then the required instrument data  (Ge single/multiple events, Ge detector count rates, 
ACS veto rates) are decoded from the telemetry packets and put into a time-sorted queue, which is large enough to hold  
several minutes of data. 

The ACS detection program ({\tt ibas\_acsmonitor}), which has already been delivered to ISDC, is explained in detail in (von Kienlin 2003)	.
The SPI-FoV software, which is comprising the two modules {\tt ibas\_smonitor\_img} and {\tt ibas\_smonitor\_rate}, is still under 
development and the delivery of the first version is expected soon.
From the telemetry packets the information on the $\gamma$-ray interaction time, the detector ID, and energy channel are extracted.
Up to now single and multiple events are analysed. This is sufficient to  do burst imaging in the instrument system, the sky coordinates
can be computed subsequently from the pointing information. 
The energy channel is used to perform a basic energy selection, 
and the data will be accumulated for time intervals 
optimized for the detection of long and short bursts.  
The parameters should be as similar as possible to those for IBIS.  
They include:  accumulation time, significance level, and energy channel range.


\begin{references}

\reference
von Kienlin, et al.\ 2003 in SPIE Conf.\ Proc., 4851, X-ray and Gamma-ray Telescopes and Instruments for Astronomy, in press 

\reference
Mereghetti, et al.\ 2001 in Proc., Gamma-Ray Bursts in the Afterglow Era, ed. E.~Costa, F.~Frontera, \& J.~Hjorth
(Berlin Heidelberg: Springer), 363

\reference
von Kienlin, et al.\ 2001 in Proc., Gamma-Ray Bursts in the Afterglow Era, ed. E.~Costa, F.~Frontera, \& J.~Hjorth
(Berlin Heidelberg: Springer), 427

\reference
Skinner, G., et al.\  1997 in ESA Conf. Proc., 382, 2$^{nd}$ INTEGRAL Workshop, ed. C.~Winkler, T.~Courvoisier, 
\& Ph.~Durouchoux (ESA), 487 

\end{references}
\end{document}